\documentclass[aps,pra,showpacs,onecolumn,superscriptaddress]{revtex4}
\usepackage{array}
\usepackage{booktabs}
\usepackage{tabu}
\usepackage{dcolumn}
\usepackage{amsmath}
\usepackage{amsfonts}
\usepackage{float}
\usepackage{amssymb}
\usepackage{graphicx,color}
\usepackage[colorlinks={true}]{hyperref}
\usepackage{graphicx}
\usepackage{subfigure}
\usepackage{graphicx}% Include figure files
\usepackage{dcolumn}% Align table columns on decimal point
\usepackage{bm}% bold math
\usepackage[T1]{fontenc}

\def\be{\begin{equation}}
  \def\ee{\end{equation}}
\def\bea{\begin{eqnarray}}
\def\eea{\end{eqnarray}}
\def\f{\frac}
\def\n{\nonumber}
\def\l{\label}
\def\p{\phi}
\def\o{\over}
\def\R{\rho}
\def\pa{\partial}
\def\om{\omega}
\def\na{\nabla}
\def\P{\Phi}
\begin{document}

\title{Control of quantum memory assisted entropic uncertainty lower bound for topological qubits in open quantum system through environment}
\author{S. Haseli}
\email{soroush.haseli@uut.ac.ir}
\affiliation{Faculty of Physics, Urmia University of Technology, Urmia, Iran.}
\author{H. Dolatkhah}
\affiliation{
Department of Physics, University of Kurdistan, P.O.Box 66177-15175, Sanandaj, Iran}

\author{H. Rangani Jahromi}
\affiliation{Physics Department, Faculty of Sciences, Jahrom University, P.B. 7413188941, Jahrom, Iran}

\author{S. Salimi}
\author{A. S. Khorashad}
\affiliation{
Department of Physics, University of Kurdistan, P.O.Box 66177-15175, Sanandaj, Iran}
\date{\today}% It is always \today, today,

\def\be{\begin{equation}}
  \def\ee{\end{equation}}
\def\bea{\begin{eqnarray}}
\def\eea{\end{eqnarray}}
\def\f{\frac}
\def\n{\nonumber}
\def\l{\label}
\def\p{\phi}
\def\o{\over}
\def\R{\rho}
\def\pa{\partial}
\def\om{\omega}
\def\na{\nabla}
\def\P{\Phi}
%\nofiles

%=============================================================%
%=============================================================%
%============== Abstract =======================================%
%=============================================================%
%=============================================================%
\begin{abstract}
The uncertainty principle is one of the most important issues that clarify the distinction between classical  and quantum theory. This principle sets a bound on our ability to predict  the  measurement outcome of two incompatible observables precisely. Uncertainty principle can be formulated via Shannon entropies of the  probability distributions of measurement outcome of the two observables. It has  shown that the entopic uncertainty bound can be improved by considering an additional particle as the quantum memory $B$ which has correlation with the measured particle $A$. In this work we consider the memory assisted entropic uncertainty for the case in which the quantum memory and measured particle are topological qubits. In our scenario the topological quantum memory $B$, is considered as an open quantum system which interacts  with its surrounding. The motivation for this model is associated  with the fact that the basis of the memory-assisted entropic uncertainty relation is constructed on the correlation between quantum memory $B$ and measured particle $A$. In the sense that, Bob who holds the quantum memory $B$  can  predict Alice's measurement results on particle $A$ more accurately, when the amount of correlation between $A$ and $B$ is great.   Here, we want to find the influence of environmental effects on uncertainty bound while the quantum memory interacts with its surrounding. In this work we will consider Ohmic-like    Fermionic and Bosonic  environment. We have also investigate the effect of the    Fermionic and Bosonic environment on the lower bounds of the amount of the key that can be extracted per state by Alice and Bob for quantum key distribution protocols. 
\end{abstract}
%\pacs{04.50}
%\keywords{keyword.}%Use showkeys class option if keyword
                              %display desired
\maketitle

%%%%%%%%%%%%%%%%%%%%%%%%%%%%%%%%%%%%%%%%%%%%%%%%%%%%%%%%%%%%%%%%%%%%%%%%%%%%
%%%%%%%%%%%%%%%%%%%%%%%%%%%%%%%%%%%%%%%%%%%%%%%%%%%%%%%%%%%%%%%%%%%%%%%%%%%%
%%%%%%%%%%%%%%%%%%%%%%%%%%%%%%%%%%%%%%%%%%%%%%%%%%%%%%%%%%%%%%%%%%%%%%%%%%%%
%%%%%%%%%%%%%%%%%%%%%%%%%%%%%%%%%%%%%%%%%%%%%%%%%%%%%%%%%%%%%%%%%%%%%%%%%%%%
%============  Sec.I (Introduction)  =======================================
%%%%%%%%%%%%%%%%%%%%%%%%%%%%%%%%%%%%%%%%%%%%%%%%%%%%%%%%%%%%%%%%%%%%%%%%%%%%
%%%%%%%%%%%%%%%%%%%%%%%%%%%%%%%%%%%%%%%%%%%%%%%%%%%%%%%%%%%%%%%%%%%%%%%%%%%%
%%%%%%%%%%%%%%%%%%%%%%%%%%%%%%%%%%%%%%%%%%%%%%%%%%%%%%%%%%%%%%%%%%%%%%%%%%%%
%%%%%%%%%%%%%%%%%%%%%%%%%%%%%%%%%%%%%%%%%%%%%%%%%%%%%%%%%%%%%%%%%%%%%%%%%%%%
\section{Introduction}	%) A SECTION HEADING
Uncertainty principle is one of the most important concept in quantum theory. Heisenberg's uncertainty relation represent the distinction between quantum theory and classical theory \cite{Heisenberg}. This relation sets a bound on our ability for precise prediction of the  measurement outcome of two incompatible observable  on a quantum system. The uncertainty principle is expressed in various form. One of the most important form of this principle was provided by Robertson \cite{Robertson} and Schrodinger \cite{Schrodinger}. According their results for any arbitrary pairs of noncommuting observables $\hat{Q}$ and $\hat{R}$, we have
\begin{equation}\label{hayzen}
\Delta \hat{Q} \Delta \hat{R} \geq \frac{1}{2} \vert \langle [ \hat{Q}, \hat{R} ]\rangle \vert,
\end{equation}
 where $\Delta \hat{Q} = \sqrt{ \langle \psi \vert \hat{Q}^{2} \vert \psi \rangle - \langle \psi \vert \hat{Q} \vert \psi \rangle^{2}}$ and $\Delta \hat{R} = \sqrt{ \langle \psi \vert \hat{R}^{2} \vert \psi \rangle - \langle \psi \vert \hat{R} \vert \psi \rangle^{2}}$are the standard deviation of the associated observable $\hat{Q}$ ($\hat{R}$) and $\left[ \hat{Q},\hat{R} \right] = \hat{Q}~\hat{R}-\hat{R}~\hat{Q}$.  It is a more efficient way to construct the uncertainty relation in terms of  Shannon entropies of the  probability distributions of measurement outcome of the two observables. The First entropic uncertainty relation was conjectured by Deutsch \cite{Deutsch}. Deutsch's entropic uncertainty relation was improved by Kraus \cite{Kraus} and then it was proved by Massen and Uffink \cite{Maassen}. They show that for any arbitrary pairs of observables $\hat{Q}$ and $\hat{R}$ with associated eigenbases $\vert q_i \rangle$ and $\vert r_i \rangle$ respectively, the entropic uncertainty can be written as
\begin{equation}\label{Massen}
H(\hat{Q})+H(\hat{R}) \geq \log_{2}\frac{1}{c},
\end{equation} 
where $H(\hat{X})=-\sum_{x} p_{x} \log_{2} p_{x}$ is the Shannon entropy of the measured observable $\hat{X} \in \lbrace \hat{Q}, \hat{R}\rbrace$, $p_x$  is the probability distributions of measurement outcome and $c=\max_{\lbrace i,j\rbrace} \vert \langle q_i \vert r_j \rangle \vert^{2}$ stands for the complementarity between the observables. Eq. \ref{Massen} can be written in general form 
\begin{equation}
H(\hat{Q})+H(\hat{R}) \geq \log_2 \frac{1}{c}+S(\hat{\rho}),
\end{equation} 
where $\hat{\rho}$ is the density matrix of measured particle and $S(\hat{\rho})=-tr(\hat{\rho} \log_2 \hat{\rho})$ is the von Neumann entropy.
The uncertainty principle  can be expressed by an interesting game between two player Alice and Bob. At the beginning of the game, Bob prepares a particle in a quantum state $\rho_A$ and sends it to Alice. In second step, they reach an agreement on measurement of two observables $\hat{Q}$ and $\hat{R}$ which is performed by Alice on her particle. Alice does measurement on her particle, and declares her choice of the measurement to Bob who wants to minimize his uncertainty about Alice's measurement outcome. If he  guesses the result of measurement accurately, he will win the game. The minimum of Bob's uncertainty about
Alice's measurement outcome is bounded by Eq. \ref{Massen}. However, when Bob prepares a correlated
bipartite state $\rho_{AB}$ and sends one part to Alice and Keeps other part as a quantum
memory by himself, he will guess the Alice's measurement outcome with a better accuracy. Entropic uncertainty relation in the existence of quantum memory is introduced by Berta et al. In Ref.\cite{Berta}, Berta et al. provide a case in which there exist a quantum memory $B$ which has correlation with measured particle $A$. Their results show that  the uncertainty of Bob about the Alice's measurement can be described by 
 \begin{equation}\label{berta}
S(\hat{Q} \vert B)+ S(\hat{R} \vert B) \geq \log_2 \frac{1}{c} + S(A \vert B),
\end{equation}
which is known as memory assisted entropic uncertainty relation, where $S(\hat{X} \vert B)=S(\rho^{XB})-S(\rho^{B})$ is  the conditional von Neumann entropies of the post measurement states 
\begin{equation}
\rho^{XB}=\sum_{i} (\vert x_i \rangle \langle x_i \vert \otimes \mathbb{I})\rho^{AB} (\vert x_i \rangle \langle x_i \vert\otimes \mathbb{I}),
\end{equation}
where $\lbrace \vert x_i \rangle\rbrace$'s are the eigenstates of the observable $\hat{X}$, and $\mathbb{I}$ is the identity operator. In the following, we will call the entropic uncertainty lower bound in Eq. \ref{berta} as Berta bound $U_{B}$.

So far, much effort has been made for tightening entropic uncertainty lower bound \cite{Pati,Pramanik,Coles,Liu,Zhang,Pramanik1,Adabi,Riccardi,Adabi1,Dolatkhah,Huang1}. In Ref. \cite{Adabi}, the aouthors have provided another bound for entropic uncertainty relation in the presence of quantum memory. They apply the same strategy to the uncertainty game in the presence of quantum memory. Based on their results, Bob's uncertainty about both $\hat{Q}$ and $\hat{R}$ measurement outcome satisfy \cite{Adabi}
\begin{eqnarray}
S(\hat{Q}\vert B)+S(R\vert B)&=&\\ \nonumber
&=&H(\hat{Q})-I(\hat{Q};B)+H(\hat{R})-I(\hat{R};B)\\ \nonumber
&\geq&\log_2 \frac{1}{c}+S(A)-[I(\hat{Q};B)+I(\hat{R};B)]\\ \nonumber
&=&\log_2 \frac{1}{c} + S(A\vert B)+ \\ \nonumber
&+& \lbrace I(A;B)- [I(\hat{Q};B)+I(\hat{R};B)]\rbrace,
\end{eqnarray}
where Eq. \ref{Massen} and $S(A)=S(A \vert B)+ I(A;B)$ are used in the second and  last
line respectively. So the entropic uncertainty relation can be rewritten as 
\begin{equation}\label{concen}
S(\hat{Q}\vert B)+S(\hat{R}\vert B) \geq \log_2 \frac{1}{c} + S(A \vert B)+ \max \lbrace 0, \delta\rbrace,
\end{equation}
where 
\begin{equation}
\delta = I(A;B)-(I(\hat{Q};B)+I(\hat{R};B)).
\end{equation}
In Eq. \ref{concen}, the uncertainties $S(\hat{Q} \vert B)$ and $S(\hat{R} \vert B)$ have lower bounded by an additional term in comparison  with Berta's uncertainty relation in Eq. \ref{berta}. If Alice measures $\hat{X} \in \lbrace \hat{Q}, \hat{R} \rbrace$, the $x$-th outcome with probability $p_x= tr_{AB}(\Pi_x^{A} \rho^{AB} \Pi_x^{A}) $ is obtained and the state of the Bob's quantum system will turn into the corresponding state $\rho_x^{B}=\frac{tr_{A}(\Pi_x^{A} \rho^{AB} \Pi_x^{A})}{p_x}$. So 
\begin{equation}
I(\hat{X};B)=S(\rho^{b})-\sum_{x}p_{x}S(\rho_{x}^{B}),
\end{equation}
is called  Holevo quantity and it is equal to the upper bound of the Bob accessible information about the outcome of Alice's measurement . In the following, we will call the entropic uncertainty lower bound in Eq. \ref{concen} as Adabi's bound $U_{A}$.   
Entropic uncertainty relations have a wide range of applications such as entanglement detection \cite{Partovi,Huang,Prevedel,Chuan-Feng} and quantum
cryptography \cite{Tomamichel,Ng}.  The security of quantum key distribution protocols can be confirmed by  the entropic uncertainty relations \cite{Ekert,Renes}. Note that the lower bound of  the uncertainty relation is directly connected with the quantum secret key (QSK)
rate. In Ref. \cite{Devetak}, it has been shown that the amount of key  that can be extracted  by Alice and Bob $K$ is lower bounded by $S(\hat{R} \vert E) - S(\hat{R} \vert B)$, where  the eavesdropper (Eve) prepares a quantum state $\rho_{ABE}$ and distributes the parts $A$
and $B$ to Alice and Bob respectively and keeps $E$. In Ref. \cite{Berta2},  Coles et al.  reconstruct their result in Eq. \ref{berta} as $ S(\hat{R} \vert E)+S(\hat{Q} \vert B) \geq \log_2 1/c$ . Based on their findings the lower bound on the QSK rate  can be written as
\begin{equation}\label{keyrateb}
K \geq \log_2 \frac{1}{c} - S(\hat{R} \vert B)-S(\hat{Q} \vert B),
\end{equation} 
In the following, we will call the quantum secret key QSK rate lower bound in Eq. \ref{keyrateb} as Berta's QSK rate bound $K_{B}$. 

From Adabi's entropic uncertainty relation in Eq. \ref{concen}, the new lower bound on the QSK rate can be obtained as 
\begin{equation}\label{keyrateH}
K \geq \log_2 \frac{1}{c} +\max \lbrace 0, \delta\rbrace- S(\hat{R} \vert B)-S(\hat{Q} \vert B),
\end{equation} 
(see \ref{appendixA} for more details). In Eq. \ref{keyrateH},  QSK rate has lower bounded by an additional term in comparison with Eq. \ref{keyrateb}.  In the following, we will call the QSK rate  lower bound in Eq. \ref{keyrateH} as Adabi's QSK rate bound  $K_{O}$.  From Eqs. \ref{keyrateb} and \ref{keyrateH}, It is observed that $K_{O}$  is tighter than $K_B$. 
In a realistic regime, it is impossible to isolate a quantum system from its surroundings subjected to information loss in the form of dissipation and decoherence. Thus, it is logical to expect that the entropic uncertainty relation can be affected by the environmental factor \cite{Zhang1,Zou,Karpat,Wang,Wang1,Zhang2,Huang2,Huang3,Wang2,Zhang3,Chen,Haseli3}. One can also  reduce the entropic uncertainty lower bound in the dissipative environment by using quantum weak measurements \cite{Haseli2,Zhang4}. Actually environmental noise can also decrease the lower bound of the QSK rate. 
Here we study the entropic uncertainty lower bound for topological qubits in the context of open quantum systems . In the sense that we consider topological qubits as the quantum memory and measured particle in our  uncertainty game such that the  topological  quantum memory interacts with its surrounding.  In this work we study the dynamics of entropic uncertainty lower bound in which  the topological quantum memory  is  coupled to the    Fermionic/Bosonic Ohmic-like environments.  Here we consider both Berta and Adabi's lower bound and compare these two with each other for topological qubit which interacts with environment.   The work is organized as follows. In Sec. \ref{sec2}, we will review the dynamics of topological qubits when they interact with    Fermionic and Bosonic environment with Ohmic-like spectral density. In Sec. \ref{sec3}, we provide our model for the memory-assisted entropic uncertainty relation in the context of open quantum systems. We  we give an example and compare the dynamics of uncertainty bound for topological qubit in different Ohmic-like environment. The manuscript closes with results
and conclusion in Sec. \ref{sec4}.
\section{ The dynamics of topological qubits }\label{sec2}
Each topological qubit is consist of two Majorana modes of a 1D Kitaev's chain which can be spatially separated. The Majorana modes are generated at the two ends of a quantum wire. They are shown by $\gamma_{a}$,$a \in \lbrace1, 2\rbrace$ where 
\begin{equation}
\gamma_{a}^{\dag}=\gamma_{a}, \quad \lbrace \gamma_{a} , \gamma_{b}\rbrace=2\delta_{ab}.
\end{equation}

These two Majorana modes interact with its surrounding in an incoherent form which leads to decoherence  of the topological qubit. The Hamiltonian of considered system can be described as
\begin{equation}
\hat{H}=\hat{H}_{\mathcal{S}}+\hat{H}_{\mathcal{E}}+\hat{H}_{int},
\end{equation}
where $\hat{H}_{\mathcal{S}}$ is the Hamiltonian of topological qubit, $\hat{H}_{\mathcal{E}}$ is the Hamiltonian of the environment and $\hat{H}_{int}$ stands for interaction between topological qubit and environment, which reads
\begin{equation}
\hat{H}_{int}=B_{1}\gamma_{1}\mathcal{O}_{1}+B_{2}\gamma_{2}\mathcal{O}_{2},
\end{equation}
where $B_{1(2)}$ is real coupling constant and $\mathcal{O}_{1(2)}$ is composite operator of electron creation $\psi_{a}^{\dag}$ and annihilation $\psi_{a}$ operator. From the Hermitian condition $\hat{H}_{int}^{\dag} =\hat{H}_{int}$ one can conclude that
\begin{equation}
\mathcal{O}_{a}^{\dag}=-\mathcal{O}_{a}. 
\end{equation} 
In the case of interaction with    Fermionic environment, Majorana modes are located at the two ends of a quantum wire with strong spin-orbit interaction. They are placed over a s-wave superconductor and driven by external magnetic fields $B$, which is applied along quantum wire direction. Each Majorana mode is coupled to a metallic nanowire by a tunnel junction with tunneling strength $B_i$ controllable by an
external gate voltage. Schematic diagram for this type of interaction is shown in Fig.(\ref{Fig1}) for the quantum memory part owned by Bob. 

In the case of interaction with Bosonic environment, Majorana modes are generated at the two ends of a quantum ring with a small gap in between.  The two Majorana modes 
has local interaction with some environmental Bosonic operator. The frequency dependence in the Bosonic environment  can be produced by an external time-dependent magnetic flux $\Phi$ which is flow through quantum ring.  Schematic diagram for this type of interaction is shown in Fig.(\ref{Fig2}) for the quantum memory part owned by Bob. 

It is worth noting that for both    Fermionic and Bosonic interaction, the environment has Ohmic-like environmental spectral density i.e. $J(\omega)\propto \omega^{s}$. The environment is known as Ohmic for $s=1$, super-Ohmic for $s>1$ and sub-Ohmic for $s<1$.

Before the interaction the state of single topological qubit (consist of two Majorana modes $\gamma_1$ and $\gamma_2$) is expand by known basis  $\vert 0 \rangle$ and $\vert 1 \rangle$ respectively. They are connected to each other by  
\begin{equation}
\frac{1}{2}(\gamma_{1}-i \gamma_2)\vert 0 \rangle=\vert 1 \rangle, \quad \frac{1}{2}(\gamma_1 +i\gamma_2)\vert 1 \rangle=\vert 0 \rangle.
\end{equation}
 It can be chosen following representation for $\gamma_{1(2)}$ 
\begin{equation}
\gamma_1=\sigma_1, \quad \gamma_{2}=\sigma_2, \quad i\gamma_{1}\gamma_{2}=\sigma_3,
\end{equation} 
where $\sigma_{i}$'s ($i=1,2,3$) are the Pauli matrices. Here the initial state $\rho_{0}$ of total system ($\mathcal{S}+\mathcal{E}$) is assumed to be uncorrelated i.e. $\rho_{0}=\rho_{\mathcal{S}}(0)\otimes \rho_{\mathcal{E}}$, where $\rho_{\mathcal{S}}(0)=\sum_{i,j=0}^{1}\rho_{ij}\vert i \rangle \langle j\vert$. In the case of    Fermionic environment, one can find the reduced density matrix of topological qubit at time $t$ as follows (see Refs. \cite{Ho} for details) 
\begin{equation}
\rho_{\mathcal{S}}^{F}(t)=\frac{1}{2}
  \left(   {\begin{array}{cc}
   1+(2 \rho_{00}-1)\alpha^{2}(t) & 2 \rho_{01}\alpha(t) \\
  2 \rho_{10}\alpha(t) & 1+(2\rho_{11}-1)\alpha^{2}(t) \\
  \end{array} } \right),
\end{equation}
while for the case of Bosonic environment it is obtained as
\begin{equation}
\rho_{\mathcal{S}}^{B}(t)=
  \left(   {\begin{array}{cc}
   \rho_{00}&  \rho_{01}\alpha(t) \\
   \rho_{10}\alpha(t) & \rho_{11} \\
  \end{array} } \right),
\end{equation}
with
\begin{equation}\label{dec}
\alpha(t)=e^{-2 B^{2}\vert \beta_{F,B} \vert \mathcal{I}_{s}(t)},
\end{equation}
and
\begin{equation}
\mathcal{I}_{s}(t)=
\begin{cases}
  2\Gamma_{0}^{s-1}\Gamma(\frac{s-1}{2})\left( 1-{}_{1}F_{1}(\frac{s-1}{2};\frac{1}{2};\frac{-\Gamma_{0}^{2}t^{2}}{4})\right)  & s \neq 1;   \\
 \frac{\Gamma_{0}^{2}t^{2}}{2} {}_{2}F_{2}\left(\lbrace1,1\rbrace ; \lbrace\frac{3}{2},2\rbrace ; \frac{-\Gamma_{0}^{2}t^{2}}{4} \right) & s=1,
 \end{cases}
\label{eq4}
\end{equation}
here $\Gamma_{0}$ indicate the environmental frequency cutoff, $\Gamma(z)$ is the Gamma function and ${}_{i}F_{j}$ is the generalized Hypergeometric function. In Eq. \ref{dec}, the $\beta_{F,B}$ are the time-independent overall coefficients which are given by 
\begin{equation}
\beta_F=\frac{-4 \pi}{\Gamma(\frac{s+1}{2})}(\Gamma_{0})^{-(s+1)}
\end{equation}
for    Fermionic environment and
\begin{equation}
\beta_{B}=
 \begin{cases}

  -\frac{N_{sc}^{2}\Gamma(3-\Delta)\epsilon^{2(\Delta-4)}}{4 \pi^{2} \Gamma(\Delta-2)2^{2\Delta-5}}\sin \pi \Delta  & 2 < \Delta \notin  \mathbf{N}  \\
 -\frac{N_{sc}^{2}\epsilon^{2(\Delta-4)}}{4 \pi(\Delta-3)!^{2}2^{2\Delta-5}} & 2 \leq \Delta \in \mathbf{N},
 \end{cases}
\label{eq4}
\end{equation}
for Bosonic environment. $N_{sc}$
is the number of degrees of freedom of the dual conformal field theory,  $\epsilon$ is the $UV$ cutoff of the length scale and $\Delta=(s+4)/2$ is conformal dimension.
\section{The dynamical model for memory-assisted entropic uncertainty relation}\label{sec3}
In this section we introduce our model to study the dynamics of enropic  uncertainty lower bound for two topological qubit which has shared between Alice and Bob. Bob prepares the correlated two topological qubit $\rho_{AB}$. The Hilbert space of two topological qubit (consist of four Majorana modes $\gamma_1$, $\gamma_2$, $\gamma_{3}$ and $\gamma_{4}$) is spanned by known basis  $\vert 00 \rangle$, $\vert 01 \rangle$, $\vert 10 \rangle$ and $\vert 11 \rangle$. They are connected to each other by 
\begin{eqnarray}
\vert 10 \rangle&=&\frac{1}{2}(\gamma_{1}-i\gamma_2)\vert 00 \rangle, \nonumber \\
\vert 01 \rangle&=&\frac{1}{2}(\gamma_{3}-i\gamma_4)\vert 00 \rangle, \nonumber \\
\vert 11 \rangle&=&\frac{1}{4}(\gamma_{1}-i\gamma_2)(\gamma_{3}-i\gamma_4) \vert 00 \rangle,
\end{eqnarray}
Due to the fact that the Majorana fermions obey the Clifford algebra one can choose 
\begin{eqnarray}
\gamma_{1}&=&\mathbb{I}\otimes\sigma_1, \quad \gamma_{2}=\mathbb{I}\otimes \sigma_{2}, \nonumber \\
\gamma_{3}&=&\sigma_{3}\otimes\sigma_{1}, \quad \gamma_{4}=\sigma_{3}\otimes \sigma_{3}.
\end{eqnarray}
\begin{figure}[!] 
\centerline{\includegraphics[scale=0.6]{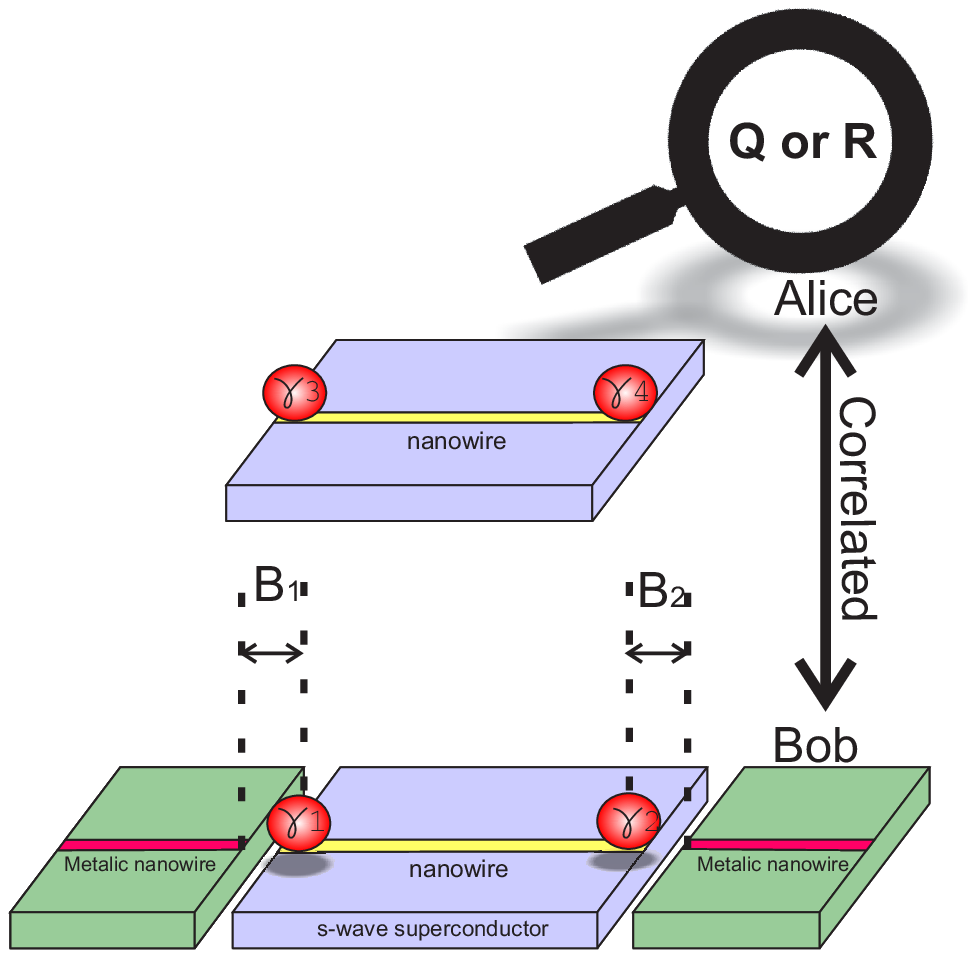}}
%\vspace*{8pt}
\caption{Schematic representation of our setting where topological quantum memory consists of a two Majorana modes interacts with    Fermionic environment. While Alice performs measurement ($\hat{Q}$ or $\hat{R}$) on her particle, and declares her choice of the measurement to
Bob. We set $B_1=B_2=B.$}\label{Fig1}
\end{figure}
 
\begin{figure}[!] 
\centerline{\includegraphics[scale=0.35]{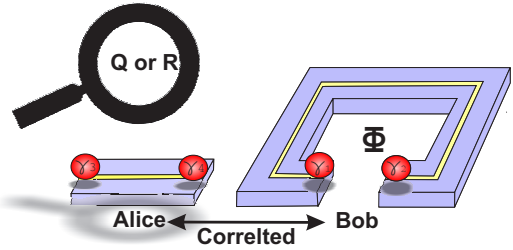}}
%\vspace*{8pt}
\caption{Schematic representation of our setting where topological quantum memory consists of a two Majorana modes interacts with Bosonic environment. While Alice performs measurement ($\hat{Q}$ or $\hat{R}$) on her particle, and declares her choice of the measurement to
Bob. }\label{Fig2}
\end{figure}
After preparing two topological qubit state by Bob, he sends one part to Alice and keeps other as a topological quantum memory. In our scenario the topological quantum memory  $B$  is an open quantum  system. So, one can show the evolution of the quantum memory  by local dynamical map $\Lambda_t$, such that the state of the two topological quantum system during evolution can be written as 
\begin{equation}\label{dynamics}
\rho_{AB_t}=(\mathbb{I}\otimes \Lambda_t)\rho_{AB}.
\end{equation}
 Next, Alice and Bob reach an agreement on measurement of two observables  which is performed by Alice on her particle. Alice does measurement on her particle, and declares her choice of the measurement to Bob who wants to minimize his uncertainty about Alice's measurement outcome. The motivation for choosing this model is related to the fact that the structure of the memory-assisted entropic uncertainty relation in Eq.(\ref{berta}), is based on the correlation between quantum memory $B$ and measured particle. Thus we want to find the usefulness and relevance of environmental effects on uncertainty bound while the quantum memory interacts with its surrounding. ِDue to the interaction of quantum memory $B$ with environment the correlation between $A$ and $B$ will decrease and so the entropic uncertanty lower bound increases while quantum secret key rate bound decreases.  Schematic representation of our setting for   Fermionic and Bosonic environment is sketched in Figs.\ref{Fig1} and \ref{Fig2}, respectively.
\subsection{Example}
\begin{figure}[h] 
\centerline{\includegraphics[scale=0.75]{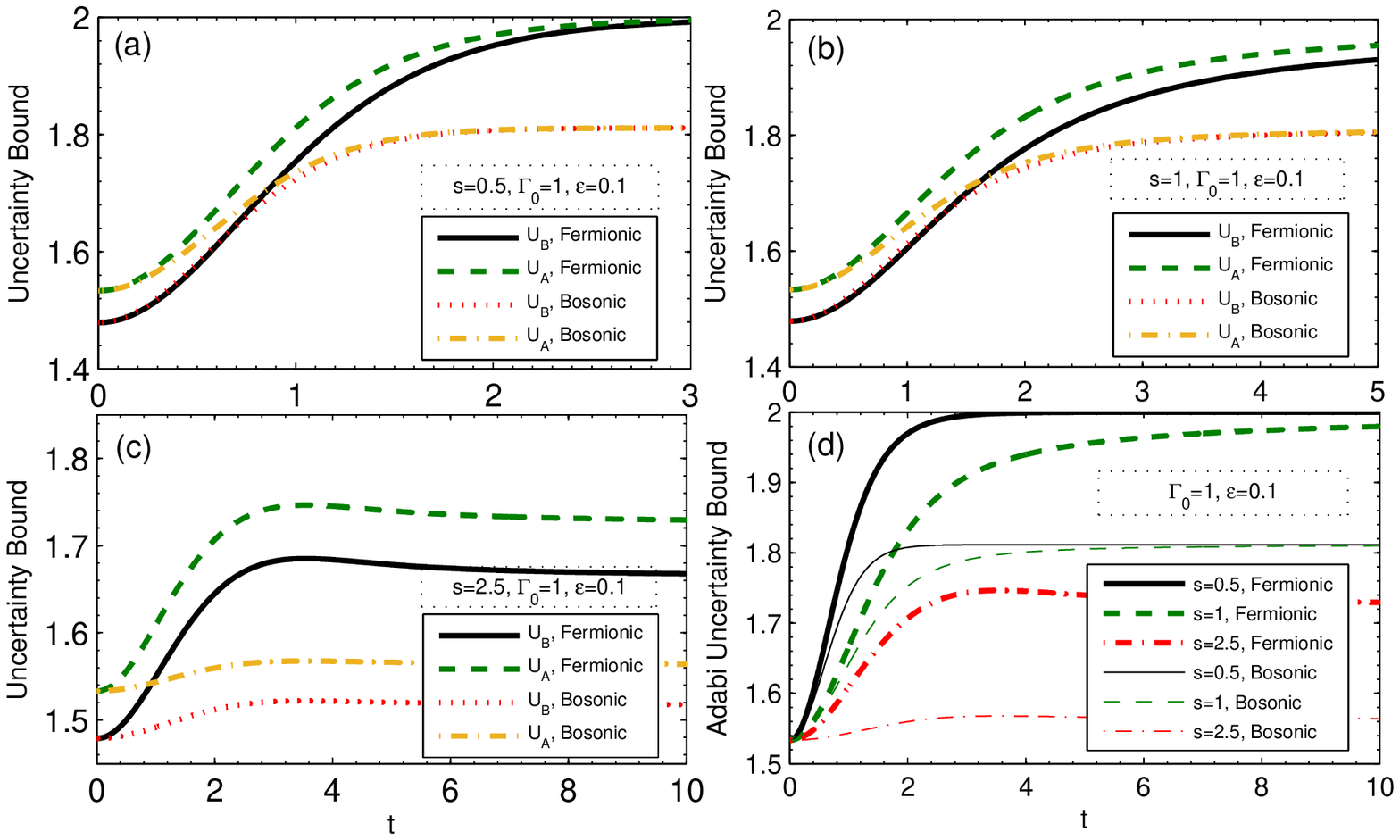}}
%\vspace*{8pt}
\caption{Entropic uncertainty lower bouns $U_{A}$ and $U_B$ as a function of time when Bob interacts with Ohmic-like    Fermionic and Bosonic environment. (a) Sub-Ohmic    Fermionic and Bosonic with $s=0.5$. (b) Ohmic    Fermionic and Bosonic with $s=1$. (c) Super-Ohmic    Fermionic and Bosonic with $s=2.5$. (d) $U_{A}$ as a function of time for different Ohmicity parameter, the thick (thin) lines stand for    Fermionic(Bosonic) Environment. We choose ($c_1=-0.6, c_{2}=c_3=0.5$), $B=0.1$ and $N_{sc}=1$ for all results.  }\label{Fig3}
\end{figure}
As an example, the set of two  topological qubit states with the maximally mixed marginal states (Bell-diagonal state) is considered as 
\begin{equation}\label{bell}
\rho^{AB}=\frac{1}{4}(I \otimes I + \sum_{k=1}^{3} c_{k}\sigma_k \otimes \sigma_k)
\end{equation}
where $\sigma_k$'s are Pauli matrices. This density matrix would be positive if $\vec{c}=(c_1,c_2,c_3)$ belongs to a tetrahedron which is defined by the set of vertices $(-1,-1,-1)$,$(-1,1,1)$,$(1,-1,1)$ and $(1,1,-1)$. Bob prepares  Bell-diagonal  two topological qubit $\rho_{AB}$, and shares it with Alice. Now, if quantum memory $B$ interacts with    Fermionic environment with Ohmic-like spectral density then from Eq. \ref{dynamics}, the dynamics of Bell-diagonal  two topological qubit can be derived as 
\begin{equation}\label{bell.f}
\rho^{AB_t}=\frac{1}{4}(I \otimes I + \sum_{k=1}^{3} c_{k}(t)\sigma_k \otimes \sigma_k)
\end{equation}
where 
\begin{equation}
c_1(t)=\alpha(t)c_1, \quad c_2(t)=\alpha(t)c_2, \quad c_3(t)=\alpha^{2}(t)c_3.
\end{equation}
In a similar way, when subsystem $B$ interacts with Bosonic environment with Ohmic-like spectral density, the time dependent coefficients  of evolved Bell-diagonal two topological qubit can be obtained as 
\begin{equation}
c_1(t)=\alpha(t)c_1, \quad c_2(t)=\alpha(t)c_2, \quad c_3(t)=c_3.
\end{equation}
From Eq. \ref{berta}, the dynamics of  Berta's entropic uncertainty lower bound $U_B$ is   given by
\begin{eqnarray}
U_B(t)&=&\log_2 \frac{1}{c}+S(A \vert B_t).
%&=&-\frac{1+c_1(t)+c_2(t)-c_3(t)}{4}\nonumber \\
%&\times& \log_2 \frac{1+c_1(t)+c_2(t)-c_3(t)}{4} \nonumber \\
%&-&\frac{1+c_1(t)-c_2(t)-c_3(t)}{4}\nonumber \\
%&\times&\log_2 \frac{1+c_1(t)-c_2(t)-c_3(t)}{4} \nonumber \\
%&-&\frac{1-c_1(t)+c_2(t)-c_3(t)}{4}\nonumber \\
%&\times& \log_2 \frac{1-c_1(t)+c_2(t)-c_3(t)}{4} \nonumber \\
%&-&\frac{1-c_1(t)-c_2(t)-c_3(t)}{4}\nonumber \\
%&\times & \log_2 \frac{1-c_1(t)-c_2(t)-c_3(t)}{4}.
\end{eqnarray}
As can be seen, unlike the Berta's  bound,  Adabi's bound depends on the measured observable.  
\begin{figure}[h] 
\centerline{\includegraphics[scale=0.75]{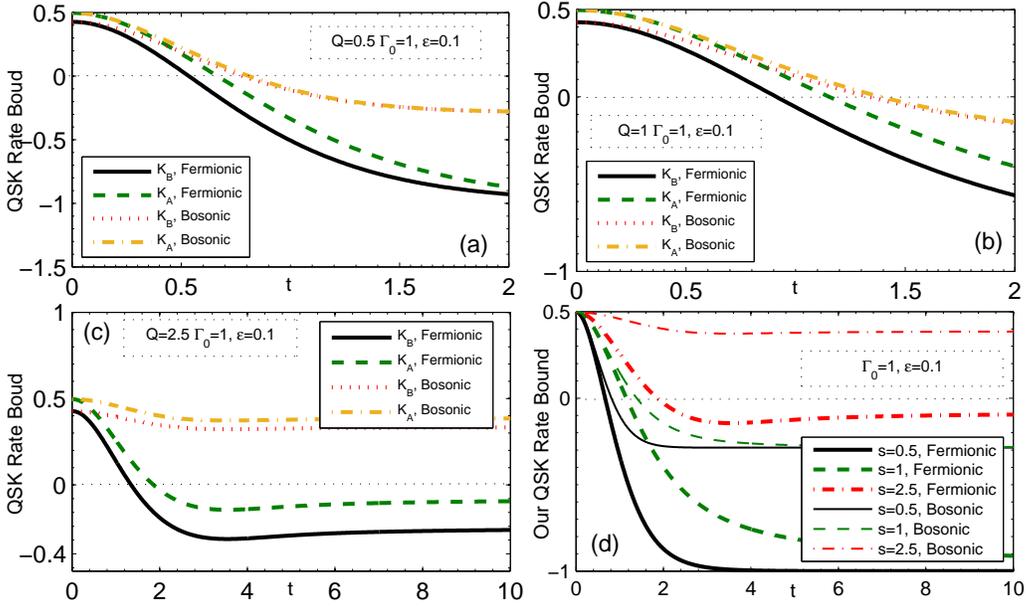}}
%\vspace*{8pt}
\caption{QSK rate bounds $K_{O}$ and $K_B$ as a function of time when Bob interacts with Ohmic-like    Fermionic and Bosonic environment. (a) Sub-Ohmic    Fermionic and Bosonic with $s=0.5$. (b) Ohmic    Fermionic and Bosonic with $s=1$. (c) Super-Ohmic    Fermionic and Bosonic with $s=2.5$. (d) $K_{O}$ as a function of time for different Ohmicity parameter, the thick (thin) lines stand for    Fermionic (Bosonic) Environment. We choose ($c_1=1, c_{2}=-c_3=-1$), $B=0.1$ and $N_{sc}=1$ for all results.}\label{Fig4}
\end{figure}

 Now we follow the straightforward strategy to obtain the dynamics of Adabi's bound. Alice can perform  her projective measurement which is represented by $P_\pm^{A}=1/2(I + \vec{n}.\vec{\sigma})$, where $\vec{n}$ is a unit vector. When Alice measures the observable $P$ on her particle, Bob's state will collapse to  $\rho_{\pm}^{B}=1/2(I+\sum_{k}n_k c_k \sigma_k)$  with probability $p_\pm=1/2$.  Given that, $\rho_B=p_+ \rho_{+}^{B}+p_- \rho_{-}^{B}$ and $S(\rho^{B})$, the time dependent Holevo quantity is obtained as\cite{Adabi}
\begin{eqnarray}\label{2}
I(P;B_t)&=&1-h(\mathcal{P}),
\end{eqnarray}
where $h(x)=- x \log_{2} x - (1-x)\log(1-x)$ and $\mathcal{P}=(1+\sqrt{(n_1c_1(t))^{2}+(n_1c_1(t))^{2}+(n_1c_1(t))^{2}})/2$. Considering the two complementary observables  $P=\hat{\sigma}_x$ and $P=\hat{\sigma}_z$ as measured observables (i.e. chosing $\vec{n}=(1,0,0)$  and $\vec{n}=(0,0,1)$ for $\hat{\sigma}_x$ and $\hat{\sigma}_z$  respectively) and from Eq. \ref{2}, the time dependent Adabi's bound is obtained as 
\begin{equation}
U_{A}=U_B+\max\lbrace 0,  I(A;B_t)-(I(\hat{\sigma}_x;B_t)+I(\hat{\sigma}_z;B_t))\rbrace.
\end{equation}
By following the similar procedure, from Berta's entropic uncertainty lower bound one can find the dynamics of the lower bound of the QSK rate as
\begin{equation}\label{keyb}
K_B(t)=\log_2 \frac{1}{c}-S(\hat{\sigma}_x \vert B_t)-S(\hat{\sigma}_z \vert B_t),
\end{equation}
and from Adabi's bound of entropic uncertainty lower bound we have
\begin{eqnarray}\label{keyh}
K_{O}(t)&=&\log_2 \frac{1}{c}-S(\hat{\sigma}_x \vert B_t)-S(\hat{\sigma}_z \vert B_t)\nonumber \\
&+&\max\lbrace 0,  I(A;B_t)-(I(\hat{\sigma}_x;B_t)+I(\hat{\sigma}_z;B_t))\rbrace.\nonumber \\
\end{eqnarray}
In Fig. \ref{Fig3}, memory assisted entropic uncertainty lower bounds for two topological qubit Bell-diagonal states with initial parameters ($c_1=-0.6, c_{2}=c_3=0.5$)  are plotted as a function of time when $B$ interacts with different types of environment. Fig. \ref{Fig3}(a), shows the dynamics of Adabi's  and Berta's entropic uncertainty lower bounds ($U_{A}$ and $U_B$ respectively) for the case that $B$ interacts with  sub-Ohmic ($s=0.5$) Fermionic and Bosonic environment. In Ref. \cite{Ho}, it has been shown that in contrast to the cases of  sub-Ohmic Fermionic environment the  correlation  in sub-Ohmic Bosonic environment does not decohere completely and it is preserved during the evolution. Due to  the dependence of entropic uncertainty lower bound on  correlation between $A$ and $B$, it is observed that entropic uncertainty lower bound reaches to its maximum value $2$ in sub-Ohmic Fermionic environment while it is suppressed in sub-Ohmic Bosonic Environment. Also, as we expect, it is observed that Adabi's entopic uncertainty lower bound is tighter than Berta's one.

In Fig. \ref{Fig3}(b), the entropic uncertainty lower bounds are considered in the case that quantum memory $B$ interacts with  Ohmic Femionic and Bosonic environment ($s=1$). As can be seen,  the results are same as the sub-Ohmic case. When $B$ interacts with  Ohmic Fermionic environment, the  correlation between $A$ and $B$ decoheres compeletely over a longer time frame in comparison with sub-Ohmic case.

In Fig. \ref{Fig3}(c), entropic uncertainty lower bounds in super-Ohmic    Fermionic and Bosonic are plotted as a function of time. As can be seen, in super-Ohmic case for Both    Fermionic and Bosonic environments correlation between $A$ and $B$ does not decohere completely. So the entopic uncertainty lower bounds do not reach to its maximum value during the evolution. From Figs. \ref{Fig3}(a), (b) and (c), one can see the Adabi's bound is tighter than Berta's bound and Bosonic environment can preserve the certainty of Bob about Alice's measurement during the quantum evolution.

The dynamics of Adabi's bound of entropic uncertainty is represented in Fig.\ref{Fig3}(d). As can be seen uncertainty lower bound is decreased by increasing Ohmicity parameter for both    Fermionic and Bosonic environments. It shows that in super-Ohmic environments correlation between $A$ and $B$ preserves during the quantum evolution. So, Bob can guess the outcome of Alice's measurement  more  accurate in super-Ohmic environments than sub-Ohmic and Ohmic environments.

Based on Adabi's and Berta's bound of entropic uncertainty relation, Berta's lower bound $K_B$ and Our's lower bound $K_{O}$ for QSK rate have been introduced in Eqs. \ref{keyrateb} and \ref{keyrateH}  respectively. In Fig. \ref{Fig4}, the lower bounds of the QSK rate  for maximally entangled two topological qubit Bell-diagonal states with initial parameters  ($c_1=1, c_{2}=-c_3=-1$) are plotted as a function of time when quantum memory $B$ interacts with different types of environments with various Ohmicity parameters. Note that the lower bounds of  the entropic uncertainty relations are directly connected with the quantum secret key rate.  

In Fig. \ref{Fig4}(a), the dynamics of Adabi's and Berta's bound of QSK rate ( \ref{keyrateH} and \ref{keyb} respectively) are plotted as a function of time. Here, the quantum memory $B$ interacts with  Fermionic and Bosonic sub-Ohmic environment $s=0.5$. It is observed that for both sub-Ohmic   Fermionic and Bosonic environments the lower bound of the amount of the key that can be extracted per state by Alice and Bob are positive just for finite initial time of the  evolution. It simply means that the sub-Ohmic Fermionic and Bosonic environment are not good enough to support quantum key distribution.

In Fig. \ref{Fig4}(b), the dynamics of QSK rates are plotted  when the quantum memory $B$ interacts with Ohmic  Fermionic and Bosonic environments. In this case, the results are same as that reported for sub-Ohmic    Fermionic and Bosonic environments. Thus, the sub-Ohmic    Fermionic and Bosonic environment are not good enough to support quantum key distribution.

 The QSK rates  for super-Ohmic    Fermionic and Bosonic environment with Ohmicity $s=2.5$ are plotted as a funcion of time in Fig.\ref{Fig4}(c).  The results are very interesting for the super-Ohmic Bosonic environment. In contrast to the super-Ohmic  Fermionic environment the QSK rates are always positive in super-Ohmic Bosonic environment. So, the super-Ohmic Bosonic environment is a suitable and desirable environment to support quantum key distribution. 
 
 In order to investigate the effect of Ohmicity parameter on QSK rate bound, our bound is plotted for different value of Ohmicity parameter in Fig.\ref{Fig4}(d). As can be seen the QSK rate bound is increased by increasing Ohmicity parameter for both    Fermionic and Bosonic environments.

\section{Conclusion}\label{sec4}
In this work, we have studied the dynamics of memory-assisted entropic uncertainty lower bounds and QSK rate bounds for the case in which the quantum memory and measured particle are topological qubits and quantum memory interacts with environment. In this work, we have considered the Berta's and Adabi's entropic uncertainty and Introduced new bound for QSK rate based on Adabi's entropic uncertainty. In \cite{Adabi}, it has been shown that Adabi's uncertainty bound is tighter than Berta's bound . 

We have considered the situation in which the topological quantum memory $B$ interacts with  Fermionic and Bosonic Ohmic-like environments. The motivation of this work stems from the fact that the foundation of memory-assisted entropic uncertainty is constructed based on the correlation which is exist between quantum memory $B$ and measured particle $A$. Due to interaction between quantum memory with surrounding the correlation betweeen quantum memory and measured particle decreases. So it is natural to expect that the uncertainty lower bound increases and QSK rate bounds decreases when quantum memory interact with environment.  In the case of two topological qubit correlation decoheres completely for sub-Ohmic and Ohmic Fermionic and Bosonic environment while it does not happen for super-Ohmic Bosonic and    Fermionic environments \cite{Ho}. We have shown that for  sub-Ohmic and Ohmic Bosonic and    Fermionic environment the uncertainty lower bounds reach to its maximum value at finite time, while it does not happen for super-Ohmic Bosonic and  Fermionic environments. So, Bob can guess the outcome of Alice's measurement  more  accurate in super-Ohmic  environment than sub-Ohmic and Ohmic environments. It has also  been shown that for both  Fermionic and Bosonic Ohmic-like environments the uncertainty bound is decreased by increasing Ohmicity parameter $s$.

We  have also shown that the QSK rate bounds for both sub-Ohmic and Ohmic    Fermionic and Bosonic environments will be negative at finite initial time. So, one can concluded that the sub-Ohmic and Ohmic environments are not good enough to support quantum key distribution protocols. That is, they are too noisy. In contrast with Ohmic and sub-Ohmic environment the QSK rate bounds are positive for super-Ohmic Bosonic environment during the interaction between quantum memory $B$ and environment. So, Super-Ohmic Bosonic environment is an ideal choice to to support quantum key distribution during time evolution. In addition, it has been shown that for both    Fermionic and Bosonic environment with different Ohmicity parameter the Adabi's QSK rate bound is tighter than Berta's rate bound. 

%% The Appendices part is started with the command \appendix;
%% appendix sections are then done as normal sections
%% \appendix

%% \section{}
%% \label{}

%% If you have bibdatabase file and want bibtex to generate the
%% bibitems, please use
%%
%%  \bibliographystyle{elsarticle-num} 
%%  \bibliography{<your bibdatabase>}

%% else use the following coding to input the bibitems directly in the
%% TeX file.
\appendix

\section{Quantum secret key rate lower bound based on Adabi's entropic uncertainty bound}\label{appendixA} 
 The main purpose of the key distribution protocol is the agreement on a shared key between two honest part (Alice and Bob) by communicating over a public channel in a way that the key is secret from any eavesdropping by the third part (Eve). The security of quantum key distribution protocols can be verified by  the entropic uncertainty relations \cite{Ekert,Renes}. The amount of key $K$ that can be extracted  by Alice and Bob  satisfy
 \begin{equation}\label{app1}
 K \geq S(\hat{R} \vert E) - S(\hat{R} \vert B).
\end{equation}  
From tripartite quantum memory uncertainty relation  
\begin{equation}
S(\hat{R} \vert E)+S(\hat{Q} \vert B) \geq \log_2 1/c \quad \rightarrow \quad S(\hat{R} \vert E)\geq \log_2 1/c -S(\hat{Q} \vert B) ,
\end{equation}
one can obtain
\begin{equation}\label{app2}
 K \geq \log_2 \frac{1}{c}- S(\hat{Q} \vert B)- S(\hat{R} \vert B).
\end{equation} 
In order to obtain the bound of quantum secret key rate from Adabi's uncertainy bound, we have to find tripartite quantum memory uncertainty relation based on Holevo quantity. For this goal we consider the following inequality for general tripartite states \cite{Coles}  
\begin{equation}\label{app3}
S(\hat{R}\vert E) \geq S(\hat{R}\vert B) -S(A \vert B) \quad \rightarrow \quad  -I(\hat{R};E) \geq -I(\hat{R};B) -S(A \vert B),
\end{equation}
the inequality convert to equality for pure tripartite states. From Eq. \ref{app3}, we have 
\begin{eqnarray}\label{app4}
S(\hat{Q}\vert B)+S(\hat{R}\vert E)&=&\\ \nonumber
&=&H(\hat{Q})-I(\hat{Q};B)+H(\hat{R})-I(\hat{R};E)\\ \nonumber
&\geq&\log_2 \frac{1}{c}+S(A)-[I(\hat{Q};B)+I(\hat{R};E)]\\ \nonumber
&=&\log_2 \frac{1}{c} + S(A\vert B)+ \\ \nonumber
&+& \lbrace I(A;B)- [I(\hat{Q};B)+I(\hat{R};E)]\rbrace,
\end{eqnarray}
from Eq. \ref{app3} we have
\begin{equation}\label{app5}
S(\hat{R} \vert E)+S(\hat{Q} \vert B) \geq \log_2 1/c + \max \lbrace 0, \delta\rbrace,
\end{equation}
where 
\begin{equation}\label{app6}
\delta = I(A;B)-(I(\hat{Q};B)+I(\hat{R};B)).
\end{equation}
By substituting Eq. \ref{app5} in to  Eq. \ref{app1},  we have following relation for the bound of  quantum secret key rate 
\begin{equation}\label{app7}
K \geq \log_2 \frac{1}{c} +\max \lbrace 0, \delta\rbrace- S(\hat{R} \vert B)-S(\hat{Q} \vert B).
\end{equation}
\bibliographystyle{unsrtnat}

\end{document}